# Corrected first law of thermodynamics for dynamical regular black holes[*]

Tianxu Huo    Chengzhou Liu[†]

Department of Physics, Shaoxing University, Shaoxing 312000, China

**Abstract:** In this work, we establish the corrected first law of thermodynamics for dynamical regular black holes on both the event horizon and apparent horizon. We find that the temperature of dynamical regular black holes derived from the traditional first law differs from that obtained through other approaches. This indicates that, similar to static cases, the first law of thermodynamics requires correction. We then derive the corrected first law of thermodynamics from the Einstein field equations. Our analysis reveals that the corrected factor originates from the fact that the $T_v^v$ component of the energy-momentum tensor depends on the black hole mass. This dependence implies that the mass of a regular black hole can no longer be directly identified as the internal energy, leading to the corrections of the first law of thermodynamics.

## 1 Introduction

In 1973, Bekenstein first noticed the striking thermodynamic analogy of black holes. He then pioneered the concept of black hole entropy as a measure of the information inside a black hole and pointed out that the black hole entropy is proportional to the horizon area: $S = A/4$ [1]. Subsequently, Hawking used the semi-classical quantum field theory to prove that black holes indeed possess a temperature: $T = \kappa/(2\pi)$, where $\kappa$ is the surface gravity [2]. The discovery of Hawking radiation laid a solid foundation for the four laws of black hole thermodynamics. In particular, for a Schwarzschild black hole, the first law of thermodynamics can be written as

$$dM = TdS, \qquad (1)$$

where $M$ denotes the mass of the black hole. Studies on other black holes such as Reissner-Nordström black hole and the Kerr-Newmann black hole revealed that they also satisfy the first law of thermodynamics.

---
[*] The work was supported by the Natural Science Foundation of Zhejiang Province of China (No. LY14A030001)
[†] Corresponding author, E-mail: czlbj20@163.com

However, when applying the first law of thermodynamics to calculate the entropy of regular black holes, one does not obtain an entropy proportional to the horizon area [3]. These black holes are characterized by having a regular core instead of a space-time singularity. Common examples of regular black holes include the Bardeen black hole, the Hayward black hole, the non-commutative black hole, and the vacuum non-singular black hole, among others. [4] and references therein provide a comprehensive overview of the thermodynamics of regular black holes. In [3], the authors pointed out that the reason of regular black holes do not satisfy the first law of thermodynamics is that the black hole mass cannot be regarded as internal energy. Based on this, the authors derived a corrected version of the first law of thermodynamics. By utilizing the corrected first law of thermodynamics, one can obtain an entropy that satisfies the Bekenstein-Hawking area law in regular black holes. In subsequent studies on the thermodynamics of regular black holes, the corrected first law of thermodynamics has been widely applied to the static black holes [5-14]. These examples include common regular black holes such as the Bardeen black hole, Hayward black hole, vacuum nonsingular black hole, and noncommutative black hole. In addition to these, we also include recently proposed nonsingular black holes. The spacetimes of these black holes are all characterized by the absence of singularities.

We know that black holes are all evolving, so it is necessary to investigate whether the first law of thermodynamics for dynamical regular black hole also requires corrected. However, for dynamical black holes, the apparent horizon and the event horizon do not coincide, leading to different viewpoints on which horizon the thermodynamics should be based. For example, in [15-17], the authors studies the entropy of a dynamical black hole using a thin-film model. By constructing the thin film outside the event horizon, they obtained an entropy proportional to the area of the event horizon. Similarly, [18-20] investigated Hawking radiation from the event horizon. On the other hand, [21] suggest that thermodynamics should be established on the apparent horizon, as it serves as the boundary of negative energy states. Considering the collapse of a spherical shell, Hiscock proposed that one-quarter of the apparent horizon area is taken as the entropy of the black hole [22]. In [23] and [24], the authors demonstrated that the first law of thermodynamics can be successfully built on the apparent horizon. They then treated the event horizon as a time-dependent perturbative hyper-surface of the apparent horizon, and also successfully established thermodynamics on the event horizon. Therefore, this paper will examine the corrected first law of thermodynamics on both horizons.

For the calculation of the temperature at the event horizon, we will use the conformal flat method [25,26]. This is because, for a dynamical black hole space-time, there are no Killing vectors, and hence, it is difficult to directly calculate the temperature at the event horizon using the surface gravity method. In addition to the conformal flat method, another approach to calculating the temperature of a dynamical black hole is the radiation back-reaction method [18]. This method first calculates the renormalized energy-momentum tensor's expectation value $\langle T_{ab} \rangle$ in the Unruh vacuum state, then examines the ingoing negative flux into the black hole, and finally determines the radiation temperature. However, this approach for studying thermal radiation from dynamical black holes is only applicable to asymptotical flat, spherically symmetric black holes and yields results with limited accuracy. Subsequently, Zhao et al. proposed the conformal flat approach, which can precisely determine the temperature and thermal spectrum of an evaporating black hole [25]. Therefore, in this study, we employ this approach to investigate the temperature at the event horizon for a general spherically symmetric dynamical regular black hole. For the apparent horizon, its surface gravity can be defined using the Kodama vector [27]. Therefore, we will directly use the surface gravity method to calculate the temperature associated to the apparent horizon.

The structure of this paper is as follows: In Sec.2, we briefly introduce the dynamical regular black holes used in this work and demonstrate their singularity-free nature by calculating their Kretschmann scalar; In Sec.3, we will calculate the special "surfaces" for a general spherically symmetric dynamical regular black hole; In Sec.4 and 5, we derive the corrected first law of thermodynamics on the event horizon and the apparent horizon, respectively; Finally, we summarize the main conclusions of this paper.

**2 Introduction to two common types of dynamical regular black holes**

In this section, we will briefly introduce the two dynamical regular black holes used as specific examples in this paper, particularly by calculating their Kretschmann scalars to clearly demonstrate that they do not possess curvature singularities at the origin. The first example is the dynamical Hayward black hole, whose line element is given by [28]

$$ds^2 = -\left[1 - \frac{2M(v)r^2}{r^3 + 2\alpha M(v)}\right]dv^2 + 2dvdr + r^2 d\theta^2 + r^2 \sin^2\theta d\varphi^2, \quad (2)$$

where $v$ is the advanced Eddington coordinate and $\alpha$ characterize quantum gravity effects having the order of squared Planck length. In the following, for brevity, we sometimes abbreviate $M(v)$ as $M$. The line element (2) is a Vaidya-type generalization of the Hayward black hole, by using the advanced Eddington coordinate. It can also be derived from the Vaidya black hole with an effective Newton constant $G' = G/(1+\alpha p^2)$, which is inspired by the generalized uncertainty principle [29]. After careful calculation, the representative geometric invariants are given by

$$\mathcal{R} = \frac{24\alpha M^2(-r^3 + 4\alpha M)}{(r^3 + 2\alpha M)^3}, \quad \mathcal{R}_{\mu\nu}\mathcal{R}^{\mu\nu} = \frac{288\alpha^2 M^4(5r^6 - 4\alpha r^3 M + 8\alpha^2 M^2)}{(r^3 + 2\alpha M)^6}, \quad (3.1)$$

$$\mathcal{R}_{\mu\nu\tau\sigma}\mathcal{R}^{\mu\nu\tau\sigma} = -\frac{4M\left[r^9 + 8(-10+\alpha)\alpha^2 M^3 - 8\alpha(-2+3\alpha)M^2 r^3 - 4(2+3\alpha)Mr^6\right]}{(r^3 + 2\alpha M)^4}.$$

(3.2)

Here, $\mathcal{R}_{\mu\nu\tau\sigma}\mathcal{R}^{\mu\nu\tau\sigma}$ is also known as the Kretschmann scalar $K$, is used to determine the existence of curvature singularities. It is evidence that the present of $\alpha$ ensures the absence of a singularity at the origin of the dynamical Hayward black hole. Setting $\alpha = 0$, Eq. (3.2) reduces to the Kretschmann scalar of the Vaidya black hole

$$K_{Vaidya} = \frac{4M(8M - r^3)}{r^6}, \qquad (4)$$

which diverges at the origin. In [30], a dynamical extension of the non-commutative Schwarzschild black hole was proposed, given by

$$ds^2 = -\left[1 - \frac{4M(v)}{\sqrt{\pi}r}\gamma\left(\frac{3}{2}, \frac{r^2}{4\vartheta}\right)\right]dv^2 + 2dvdr + r^2d\theta^2 + r^2\sin^2\theta d\varphi^2, \qquad (5)$$

where the lower incomplete gamma function is defined by

$$\gamma\left(\frac{3}{2}, \frac{r^2}{4\vartheta}\right) = \int_0^{r^2/4\vartheta} t^{1/2} e^{-t} dt. \qquad (6)$$

Here, $\vartheta$ is the non-commutative parameter with dimensions of Planck length squared, representing the smeared structure of space. In non-commutative space, geometric points that describe positions are replaced by regions with a minimum width on the order of the Planck length. Therefore, the method of defining point mass density using the Dirac delta function is no longer applicable; instead, a Gaussian distribution is used. The line element (5) gradually approaches that of a Vaidya black hole as $r$ increase. For the dynamical non-commutative black hole, we can also calculate the geometric invariants, which turn out to be

$$\mathcal{R} = -\frac{e^{-\frac{r^2}{4\vartheta}} M(r^2 - 8\vartheta)}{2\sqrt{\pi}\vartheta^{5/2}}, \quad \mathcal{R}_{\mu\nu}\mathcal{R}^{\mu\nu} = \frac{e^{-\frac{r^2}{2\vartheta}} M^2 (r^4 - 8r^2\vartheta + 32\vartheta^2)}{8\pi\vartheta^5}, \qquad (7.1)$$

$$\mathcal{R}_{\mu\nu\tau\sigma}\mathcal{R}^{\mu\nu\tau\sigma} = \frac{M}{2\pi}\left\{\frac{e^{-\frac{r^2}{2\vartheta}}}{\vartheta^3}\left(8M + \sqrt{\pi\vartheta}r^2 e^{\frac{r^2}{4\vartheta}}\right) - \frac{16}{r^3}\left(\sqrt{\pi} + \frac{4Me^{-\frac{r^2}{4\vartheta}}}{\vartheta^{3/2}}\right)\gamma\left(\frac{3}{2}, \frac{r^2}{4\vartheta}\right)\right.$$

$$\left. + \frac{256M}{r^6}\left[\gamma\left(\frac{3}{2}, \frac{r^2}{4\vartheta}\right)\right]^2\right\}. \qquad (7.2)$$

For small $r$, using the definition of the gamma function, we have

$$\gamma\left(\frac{3}{2}, \frac{r^2}{4\vartheta}\right) \approx \int_0^{r^2/4\vartheta} t^{1/2} dt = \frac{2}{3}\left(\frac{r^2}{4\vartheta}\right)^{3/2} = \frac{r^3}{12\vartheta^{3/2}}, \qquad (8)$$

therefore, the Kretschmann scalar is finite at the origin. When $r \to 0$, we have

$$\lim_{r \to 0}\left(\mathcal{R}_{\mu\nu\tau\sigma}\mathcal{R}^{\mu\nu\tau\sigma}\right) = \frac{2M\left(10M - 3\sqrt{\pi}\vartheta^{3/2}\right)}{9\pi\vartheta^3}, \tag{9}$$

which shows that the non-commutative factor $\vartheta$ ensures the absence of a curvature singularity. Additionally, when $\vartheta \to 0$, we can similarly obtain the Kretschmann scalar for the Vaidya black hole. In Fig.1, we have plotted the relationship between the Kretschmann scalar and $r$ for two types of dynamical black holes. It can be seen that there is no curvature singularity at the origin in both cases.

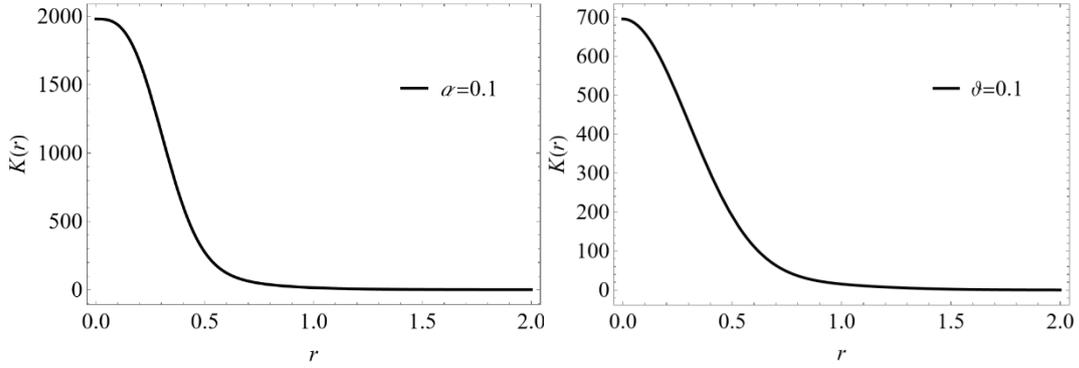

Fig.1 The Kretschmann scalar as a function of $r$: (a) for the dynamical Hayward black hole, and (b) for the dynamical non-commutative black hole. Here, we set $M(v) = 1$.

**3 Horizons of a spherically symmetric dynamical Black hole**

In order to calculate the horizon radius of a general spherically symmetric dynamical black hole, we assume its line element to be

$$ds^2 = -f(v,r)dv^2 + 2dvdr + r^2d\theta^2 + r^2\sin^2\theta d\varphi^2, \tag{10}$$

where $f(v,r) = 1 - 2m(v,r)/r$. The determinant and non-vanishing contravariant components of the metric are given by, respectively,

$$g = -r^4\sin^2\theta,$$

$$g^{vr} = g^{rv} = 1, \quad g^{rr} = f(v,r), \quad g^{\theta\theta} = \frac{1}{r^2}, \quad g^{\varphi\varphi} = \frac{1}{r^2\sin^2\theta}. \tag{11}$$

The line element (10) can also be rewritten as $ds^2 = h_{ab}dx^a dx^b + r^2 d\Omega^2$, with $x^a = (v, r)$. In general, a dynamical black hole is characterized by different special "surfaces": the time-like limit surface, the trapped horizon, the apparent horizon and the event horizon. The definition of the time-like limit surface of a dynamical black hole is the same as that in a stationary black hole: $g_{vv} = 0$, therefore, the radius of the time-like limit surface satisfies $f(v, r_{TLS}) = 0$. According to the definition of the trapped horizon: $h^{ab} \partial_a r \partial_b r = 0$, we can obtain that the radius of the trapped horizon also satisfies $f(v, r_{TH}) = 0$. The apparent horizon is the outermost marginally trapped surface and is defined as $\Theta = l^{\mu}_{;\mu} - \kappa = 0$, where $\Theta$ is the expansion of a congruence of null geodesics and $\kappa = l_{\mu;\nu} n^{\mu} l^{\nu}$ reduces to the surface gravity in the stationary case. Additionally, $n^{\mu}$ and $l^{\nu}$ are null tetrads. For the spherically symmetric dynamical black hole described by line element (10), we calculate the radius of its apparent horizon in the Appendix. Here, we directly use the result. The radius satisfies $f(v, r_{AH}) = 0$, that is $r_{AH} = m(v, r_{AH})$. We can see that for an arbitrary spherically symmetric dynamical black hole, the trapped horizon and the apparent horizon coincide. Therefore, in this paper, we denote their radius uniformly as $r_{AH}$.

Next, we proceed to calculate the radius of the event horizon. The event horizon is defined as a special hyper-surface where the norm of the normal vector is zero (while the normal vector itself is non-zero). It also referred to as a null hyper-surface, and this hyper-surface preserves the symmetries of the space-time. For the spherically symmetric space-time described by (10), the event horizon can be assumed to be of the form $F(v, r)$, with the normal vector defined as $n_{\mu} = \partial_{\mu} F(v, r)$. From the condition that the normal vector is null, we arrive at

$$n_{\mu} n^{\mu} = g^{\mu\nu} \frac{\partial F}{\partial x^{\mu}} \frac{\partial F}{\partial x^{\nu}} = 0. \tag{12}$$

Using the null surface equation $F(v, r) = 0$, we have

$$\frac{dr}{dv} \frac{\partial F}{\partial r} + \frac{\partial F}{\partial v} = 0. \tag{13}$$

Substituting Eq. (13) into Eq. (12), we find that the event horizon radius $r_H$ satisfies

$$f(v, r_H) - 2\dot{r}_H = 0. \tag{14}$$

In this paper, we donate the event horizon radius as $r_H$. Since $\dot{r}_H$ is a small quantity, it is easy to see that the event horizon and the apparent horizon are very close to each other.

**4 Thermodynamics on the event horizon**

**4.1 Temperature**

In this section, we will use the so-called conformal flat approach to calculate the temperature of dynamical regular black holes. The core idea of this approach is that, in a general static or stationary space-time, one can employ the tortoise coordinate transformation to render the space-time explicitly conformal to Minkowski space near the horizon. As a result, the dynamical equations describing particles can be reduced to their traditional form near the horizon. Subsequently, Zhao et al. applied the conformal flat approach to a dynamical black hole to calculate the temperature, yielding a more accurate result than the widely used radiation back-reaction method [25].

We start with the Klein-Gordon equation describing the motion of a scalar field with mass $m_0$

$$\frac{1}{\sqrt{-g}} \partial_\mu \left( \sqrt{-g} g^{\mu\nu} \partial_\nu \psi \right) - m_0^2 \psi = 0. \tag{15}$$

Combining with the line element (10), the Klein-Gordon equation can be explicitly written as

$$f \frac{\partial^2 \psi}{\partial r^2} + 2 \frac{\partial^2 \psi}{\partial v \partial r} + \frac{2}{r} \frac{\partial \psi}{\partial v} + \left( \frac{2f}{r} + f' \right) \frac{\partial \psi}{\partial r} \\
+ \frac{1}{r^2} \left[ \frac{1}{\sin\theta} \frac{\partial}{\partial \theta} \left( \sin\theta \frac{\partial \psi}{\partial \theta} \right) + \frac{1}{\sin^2\theta} \frac{\partial^2 \psi}{\partial \varphi^2} \right] - m_0^2 \psi = 0. \tag{16}$$

In the dynamical regular black hole space-time, we separate the variables of the wave function as follows

$$\psi = \frac{1}{r} R(v, r) Y_{lm}(\theta, \varphi). \tag{17}$$

Substituting the separated wave function $\psi$ into Eq. (16), we obtain the radical and angular components of the field equations as follows

$$f\frac{\partial^2 R}{\partial r^2} + 2\frac{\partial^2 R}{\partial v \partial r} + f'\frac{\partial R}{\partial r} - \left[\frac{f'}{r} - \frac{l(l+1)}{r^2} + m_0^2\right]R = 0, \tag{18.1}$$

$$\left[\frac{1}{\sin\theta}\frac{\partial}{\partial\theta}\left(\sin\theta\frac{\partial}{\partial\theta}\right) + \frac{1}{\sin^2\theta}\frac{\partial^2}{\partial\varphi^2} - l(l+1)\right]Y_{lm} = 0. \tag{18.2}$$

Here, $l$ is the angular quantum number of the particle. As the angular equation is not relevant to our discussion, the following analysis will focus on the radial part. Next, consider the following coordinate transformation

$$r_* = \frac{1}{2\kappa(v_0)}\ln\left|\frac{r - r_H(v)}{r_H(v_0)}\right|, \tag{19}$$

$$v_* = v - v_0,$$

where $v_0$ represents the moment of a particle when it escapes from the horizon, and $\kappa(v_0)$ is a function yet to be determined. Both $r(v_0)$ and $\kappa(v_0)$ represent the values of the corresponding quantities at the moment the particle leaves the horizon. Therefore, they can be treated as constants under the coordinate transformation. Differentiating Eq. (19), we obtain

$$dr_* = \frac{1}{2\kappa(r - r_H)}dr - \frac{\dot{r}_H}{2\kappa(r - r_H)}dv, \tag{20}$$

$$dv_* = dv.$$

Then, the transformation of the differential operator under the tortoise coordinate transformation can be further obtained as

$$\frac{\partial}{\partial r} = \frac{1}{2\kappa(r - r_H)}\frac{\partial}{\partial r_*}, \quad \frac{\partial}{\partial v} = \frac{\partial}{\partial v_*} - \frac{\dot{r}_H}{2\kappa(r - r_H)}\frac{\partial}{\partial r_*},$$

$$\frac{\partial^2}{\partial r^2} = \left[\frac{1}{2\kappa(r - r_H)}\right]^2\frac{\partial^2}{\partial r_*^2} - \frac{1}{2\kappa(r - r_H)^2}\frac{\partial}{\partial r_*}, \tag{21}$$

$$\frac{\partial^2}{\partial r \partial v} = \frac{1}{2\kappa(r - r_H)}\frac{\partial^2}{\partial r_* \partial v_*} - \frac{\dot{r}_H}{[2\kappa(r - r_H)]^2}\frac{\partial^2}{\partial r_*^2} + \frac{\dot{r}_H}{2\kappa(r - r_H)^2}\frac{\partial}{\partial r_*}.$$

With the transformation of the differential operator in hand, the radial equation of motion for the particle can be rewritten as

$$\frac{f - 2\dot{r}_H}{2\kappa(r - r_H)} \frac{\partial^2 R}{\partial r_*^2} + \left(-\frac{f - 2\dot{r}_H}{r - r_H} + f'\right) \frac{\partial R}{\partial r_*} + 2\frac{\partial^2 R}{\partial r_* \partial v_*}$$
$$- 2\kappa(r - r_H)\left[\frac{f'}{r} + \frac{l(l+1)}{r^2} + m_0^2\right] R = 0. \tag{22}$$

According to the spirit of the conformal flat approach, the coefficient of $\partial^2 R/\partial r_*^2$ should be equal to 1 near the event horizon at the moment $v_0$ when the particle escapes from the event horizon. At the same time, it is noted that

$$\lim_{\substack{r \to r_H(v_0) \\ v \to v_0}} \left[f(v,r) - 2\dot{r}_H\right] = 0, \tag{23}$$

where we have used the relation satisfied by the horizon radius, Eq. (14). Therefore, we can apply L'Hôpital's rule to evaluate the coefficient of $\partial^2 R/\partial r_*^2$

$$\lim_{\substack{r \to r_H(v_0) \\ v \to v_0}} \left[\frac{f(v,r) - 2\dot{r}_H}{2\kappa(v_0)(r - r_H)}\right] = \lim_{\substack{r \to r_H(v_0) \\ v \to v_0}} \left[\frac{f'(v,r)}{2\kappa(v_0)}\right] = 1. \tag{24}$$

Using the above expression, we can finally determine $\kappa(v)$ at any given advanced Eddington time

$$\kappa(v) = \frac{1}{2} f'(v, r_H). \tag{25}$$

Meanwhile, using L'Hôpital's role, we can also prove that

$$\lim_{\substack{r \to r_H(v_0) \\ v \to v_0}} \left[-\frac{f(v,r) - 2\dot{r}_H}{r - r_H}\right] = -\lim_{\substack{r \to r_H(v_0) \\ v \to v_0}} f'(v,r). \tag{26}$$

As a result, the coefficient of $\partial R/\partial r_*$ vanishes in the above limit. It is not difficult to show that the coefficient of $R$ also break down near the horizon. Therefore, the radial equation of motion reduces to the traditional form

$$\frac{\partial^2 R}{\partial r_*^2} + 2\frac{\partial^2 R}{\partial v_* \partial r_*} = 0. \tag{27}$$

The two linearly independent solutions of Eq. (27) are

$$R_\omega^{\text{in}} = e^{-i\omega v_*},$$
$$R_\omega^{\text{out}} = e^{-i\omega v_* + 2i\omega r_*}. \tag{28}$$

Following the approach of Damour and Ruffini [31] and Sannan [32], we can obtain the spectral distribution of the outgoing wave

$$N_\omega = \frac{\Gamma_\omega}{e^{\omega/T_H} \pm 1}, \tag{29}$$

with

$$T_H = \frac{\kappa}{2\pi}. \tag{30}$$

Here, " + " corresponds to fermions and " − " represent the bosons, $\Gamma_\omega$ is the transmission coefficient cause by gravitational field barrier outside the horizon. From Eq. (29) and Eq. (30), we can see that $\kappa$ should be identified as the surface gravity at the horizon. Substituting Eq. (25) into Eq. (30), we can obtain the temperature at the event horizon as

$$T_H = \frac{\kappa}{2\pi} = \frac{f'(v,r_H)}{4\pi} = \frac{1-2\dot{r}_H}{4\pi r_H} - \frac{m'(v,r_H)}{2\pi r_H}. \tag{31}$$

Next, we will specifically calculate the temperature at the event horizon of the two dynamically regular black holes introduced in Sec.2. For the dynamical Hayward black hole, using Eq. (14), we can obtain that its event horizon radius satisfies

$$r_H^3 - \left(\frac{2M}{1-2\dot{r}_H}\right)r_H^2 + 2\alpha M = 0, \tag{32}$$

which gives the relationship between the black hole mass and the event horizon radius

$$M = \frac{(1-2\dot{r}_H)r_H^3}{2\left[r_H^2 + (2\dot{r}_H - 1)\alpha\right]}. \tag{33}$$

Combining Eq. (31), we obtain the temperature at the event horizon as

$$T_H = \frac{\kappa}{2\pi} = \frac{(1-2\dot{r}_H)\left[r_H^2 + 3(2\dot{r}_H - 1)\alpha\right]}{4\pi r_H^3}. \tag{34}$$

Similarly, the event horizon radius of the dynamical non-commutative black hole satisfies

$$1 - \frac{4M}{\sqrt{\pi} r_H}\gamma\left(\frac{3}{2}, \frac{r_H^2}{4\vartheta}\right) - 2\dot{r}_H = 0. \tag{35}$$

The temperature is given by

$$T_H = \frac{\kappa}{2\pi} = \frac{1}{4\pi}f'(v,r_H) = \frac{(2\dot{r}_H - 1)}{16\pi r_H}\left[-4 + r_H^3 \vartheta^{-\frac{3}{2}} e^{-\frac{r_H^2}{4\vartheta}} \gamma^{-1}\left(\frac{3}{2}, \frac{r_H^2}{4\vartheta}\right)\right]. \tag{36}$$

On the other hand, the corresponding temperatures of the dynamical Hayward black hole and the dynamical non-commutative black hole given by the first law of black hole thermodynamics are respectively

$$\tilde{T}_H = \frac{\partial M}{\partial S_H} = \frac{\partial r_H}{\partial S_H}\frac{\partial M}{\partial r_H} = \frac{(1-2\dot{r}_H)\left[r_H^2 + 3(2\dot{r}_H - 1)\alpha\right]r_H}{4\pi\left[r_H^2 + (2\dot{r}_H - 1)\alpha\right]^2}, \tag{37}$$

$$\tilde{T}_H = \frac{\partial M}{\partial S_H} = \frac{2\dot{r}_H - 1}{32\sqrt{\pi}r_H}\left[r_H^3 \vartheta^{-\frac{3}{2}} e^{-\frac{r_H^2}{4\vartheta}} - 4\gamma\left(\frac{3}{2},\frac{r_H^2}{4\vartheta}\right)\right]\gamma^{-2}\left(\frac{3}{2},\frac{r_H^2}{4\vartheta}\right). \tag{38}$$

The relation $S_H = \pi r_H^2$ [23,24] was used in the derivation above. It is clearly seen that the temperature obtained from the first law of thermodynamics is different from that obtained using the conformal flat method. Similar to the case of static black holes [3,5-14], this will inspire us to search for the corrected first law of thermodynamics for dynamical black holes.

**4.2 Corrected first law of thermodynamics on event horizon**

From the above calculations, we can see that, similar to static regular black holes, dynamical regular black holes also do not satisfy the first law of thermodynamics. In the following, we will attempt to find a corrected version of the first law of thermodynamics. To this end, we rewrite the relation satisfied by the event horizon as

$$m(v,r_H) = \frac{r_H}{2}(1-\dot{r}_H). \tag{39}$$

On the other hand, the Einstein filed equations corresponding to line element (10) are

$$\frac{\partial m(v,r)}{\partial v} = 4\pi r^2 T_v^r. \tag{40.1}$$

$$\frac{\partial m(v,r)}{\partial r} = -4\pi r^2 T_v^v. \tag{40.2}$$

Using Eq. (40.2), we can rewrite the temperature as

$$T = \frac{1-2\dot{r}_H}{4\pi r_H} + 2r_H T_v^v. \tag{41}$$

Integrating Eq. (40.2), we can obtain

$$m(v,r) = M(v) + 4\pi \int_r^\infty r^2 T_v^v dr. \tag{42}$$

At the event horizon, Eq. (42) gives

$$M(v) = m(v, r_H) - 4\pi \int_{r_H}^{\infty} r^2 T_v^v dr. \tag{43}$$

Differentiating the above equation, on can obtain

$$dM(v) = d\left[\frac{r_H}{2}(1 - 2\dot{r}_H)\right] - 4\pi d\left(\int_{r_H}^{\infty} r^2 T_v^v dr\right). \tag{44}$$

For regular black holes, $T_v^v$ is a function of $M(v)$. Therefore, the differential of the second term on the right-hand side of the above equation becomes

$$-4\pi d\left(\int_{r_H}^{\infty} r^2 T_v^v dr\right) = -4\pi T_v^v r_H^2 dr_H - 4\pi \left[\int_{r_H}^{\infty} r^2 \frac{\partial T_v^v}{\partial M(v)} dr\right] dM(v). \tag{45}$$

Therefore, we can rearrange Eq. (44) as

$$\left[1 + 4\pi \int_{r_H}^{\infty} r^2 \frac{\partial T_v^v}{\partial M(v)} dr\right] dM(v) = \left[\frac{1 - 2\dot{r}_H}{4\pi r_H} + 2 r_H T_v^v\right] d\left(\frac{A_H}{4}\right). \tag{46}$$

Combining with the temperature expression (41), the above equation can be simplified to

$$\left[1 + 4\pi \int_{r_H}^{\infty} r^2 \frac{\partial T_v^v}{\partial M(v)} dr\right] dM(v) = T_H dS_H. \tag{47}$$

This is the corrected first law of thermodynamics for the dynamical regular black hole, and its form is the same as that in the case of a static black hole. The correction factor in above expression is: $1 + 4\pi \int_{r_H}^{\infty} r^2 \partial T_v^v / \partial M(v) dr$. For convenience in application, we use Eq. (42) to rewrite the corrected first law of black hole thermodynamics as

$$d\mathcal{M}(v)\big|_{r=r_H} = \left[\frac{\partial m(v,r)}{\partial M(v)}\bigg|_{r=r_H}\right] dM(v) = T_H dS_H. \tag{48}$$

We can clearly see that the relationship between $T_H$ and $\tilde{T}_H$ is given by

$$T_H = \left[\frac{\partial m(v,r)}{\partial M(v)}\bigg|_{r=r_H}\right] \tilde{T}_H. \tag{49}$$

When additional variables appear in the expression for $T_v^v(r, M(v), \alpha, \beta, \ldots)$, that is, in the presence of other fields, we can readily express the modified first law of thermodynamics as

$$d\mathcal{M}(v)\big|_{r=r_H} = \left[\frac{\partial m(v,r)}{\partial M(v)}\right]_{r=r_H} dM(v) = T_H dS_H + \Phi_H^\alpha d\alpha + \Phi_H^\beta d\beta + \cdots. \tag{50}$$

For the dynamical Hayward black holes, the correction factor to the first law of thermodynamics can be calculated as

$$F(v,r_H,\alpha) = \frac{\partial m(v,r,\alpha)}{\partial M(v)}\bigg|_{r=r_H} = \left[\frac{\partial}{\partial M(v)}\left(\frac{M(v)r^3}{r^3+2\alpha M(v)}\right)\right]_{r=r_H} = \frac{1}{r_H^4}\left[r_H^2 + (2\dot{r}_H - 1)\alpha\right]^2. \tag{51}$$

The product of $T_H$ and the correction factor $F(v,r_H,\alpha)$ exactly gives $\tilde{T}_H$. Similarly, for the dynamical non-commutative black holes, its correction factor is given by

$$F(v,r_H,\vartheta) = \frac{\partial m(v,r,\vartheta)}{\partial M(v)}\bigg|_{r=r_H} = \left\{\frac{\partial}{\partial M(v)}\left[\frac{2M(v)}{\sqrt{\pi}}\gamma\left(\frac{3}{2},\frac{r^2}{4\vartheta}\right)\right]\right\}_{r=r_H} = \frac{2}{\sqrt{\pi}}\gamma\left(\frac{3}{2},\frac{r_H^2}{4\vartheta}\right). \tag{52}$$

A direct calculation can also verify that: $T_H = F(v,r_H,\vartheta)\tilde{T}_H$, which demonstrates the validity of the corrected first law of thermodynamics.

## 5 Corrected first law of thermodynamics at apparent horizon and some remarks
## 5.1 Corrected first law of thermodynamics at apparent horizon

In a dynamical spherically symmetric space-time, the Kodama vector can be used to describe the symmetry of the space-time. When the dynamical space-time reduces to a static one, the Kodama vector also reduces to the Killing vector. The Kodama vector is defined as $K^a = -\epsilon^{ab}\nabla_b r$, where $\epsilon^{ab}$ denotes the volume form [27]. For the metric (2), the Kodama vector is given by $K^a = (\partial_v)^a$. As the Kodama vector is divergence free $\nabla_a K^a = 0$, there exists a conserved current $J^a = T^a_b K^b$, and the conserved charge is $E = -\int_\sigma J^a d\sigma_a$, which is equal to the Misner-Sharp energy. Using the Kodama vector, Hayward defined the surface gravity on the apparent horizon as $\kappa = \frac{1}{2}\nabla^a \nabla_a r$. Therefore, we can directly use the surface gravity method to calculate the temperature at the apparent horizon. In combination with the line element (2), we obtain the surface gravity associated the apparent horizon as

$$\kappa = \frac{1}{2}\nabla^a \nabla_a r \bigg|_{r=r_{AH}} = \left[\frac{1}{2\sqrt{-h}}\partial_i\left(\sqrt{-h}h^{ij}\partial_j r\right)\right]\bigg|_{r=r_{AH}}$$

$$= \left[\frac{1}{2\sqrt{-h}}\partial_r\left(\sqrt{-h}h^{rr}\right)\right]\bigg|_{r=r_{AH}} \tag{53}$$

$$= \frac{1}{2}f'(v, r_{AH}).$$

Here, $i$ and $j$ run from $0$ to $1$. The temperature derived from the surface gravity method is given by

$$T_{AH} = \frac{\kappa}{2\pi} = \frac{f'(v, r_{AH})}{4\pi} = \frac{1}{4\pi r_{AH}} - \frac{m'(v, r_{AH})}{2\pi r_{AH}}. \tag{54}$$

The last step in the above equation makes use of the condition satisfied by the radius of the apparent horizon: $r_{AH} = 2m(v, r_{AH})$. Next, we calculate the dynamical regular black holes discussed in this paper. For dynamical Hayward black holes, the relation between $M(v)$ and $r_{AH}$ is

$$M(v) = \frac{r_{AH}^3}{2(r_{AH}^2 - \alpha)}. \tag{55}$$

Combing with Eq. (54), we obtain the temperature at the apparent horizon as

$$T_{AH} = \frac{r_{AH}^2 - 3\alpha}{4\pi r_{AH}^3}. \tag{56}$$

Following the same approach, the temperature at the apparent horizon for the dynamical non-commutative black hole is given by

$$T_{AH} = \frac{1}{16\pi r_{AH}}\left[4 - r_{AH}^3 \vartheta^{-\frac{3}{2}} e^{-\frac{r_{AH}^2}{4\vartheta}} \gamma^{-1}\left(\frac{3}{2}, \frac{r_{AH}^2}{4\vartheta}\right)\right]. \tag{57}$$

Next, we can also use the first law of thermodynamics [23,24] to obtain the temperature of the apparent horizons for the two black holes, which are given by

$$\tilde{T}_{AH} = \frac{\partial M}{\partial S_{AH}} = \frac{\partial r_{AH}}{\partial S_{AH}}\frac{\partial M}{\partial r_{AH}} = \frac{r_{AH}\left(r_{AH}^2 - 3\alpha\right)}{4\pi\left(r_{AH}^2 - \alpha\right)^2}, \tag{58}$$

$$\tilde{T}_{AH} = \frac{1}{32\sqrt{\pi}r_{AH}}\left[r_{AH}^3 \vartheta^{-\frac{3}{2}} e^{-\frac{r_{AH}^2}{4\vartheta}} - 4\gamma\left(\frac{3}{2}, \frac{r_{AH}^2}{4\vartheta}\right)\right]\gamma^{-2}\left(\frac{3}{2}, \frac{r_{AH}^2}{4\vartheta}\right). \tag{59}$$

It is seen that, like the case at the event horizon, the temperatures obtained from the surface gravity and the first law of thermodynamics at the apparent horizon are also different. Now, we follow a similar procedure as in the previous section to derive the corrected first law of thermodynamics at the apparent horizon. First, using Eq. (40.2), the temperature can be expressed as

$$T_{AH} = \frac{1}{4\pi r_{AH}} + 2r_{AH}T_v^v. \tag{60}$$

At the apparent horizon, Eq. (42) gives

$$M(v) = m(v, r_{AH}) - 4\pi \int_{r_{AH}}^{\infty} r^2 T_v^v dr. \tag{61}$$

Substituting $m(v, r_{AH}) = r_{AH}/2$ and differentiating Eq. (61), we obtain

$$dM(v) = d\left(\frac{1}{2}r_{AH}\right) - 4\pi d\left(\int_{r_{AH}}^{\infty} r^2 T_v^v dr\right). \tag{62}$$

Eq. (62) finally gives the following result

$$\left[1 + 4\pi \int_{r_{AH}}^{\infty} r^2 \frac{\partial T_v^v}{\partial M(v)} dr\right] dM(v) = \left[\frac{1}{4\pi r_{AH}} + 2r_{AH}T_v^v\right] d\left(\frac{A_{AH}}{4}\right). \tag{63}$$

Combining with the temperature expression (60), Eq. (63) can be simplified to

$$\left[1 + 4\pi \int_{r_{AH}}^{\infty} r^2 \frac{\partial T_v^v}{\partial M(v)} dr\right] dM(v) = T_{AH} dS_{AH}. \tag{64}$$

Using Eq. (42), we finally obtain the corrected first law of thermodynamics at the apparent horizon as

$$d\mathcal{M}(v)\Big|_{r=r_{AH}} = \left[\frac{\partial m(v,r)}{\partial M(v)}\Big|_{r=r_{AH}}\right] dM(v) = T_{AH} dS_{AH}. \tag{65}$$

When other fields are present, Eq. (65) can be generalized as

$$d\mathcal{M}(v)\Big|_{r=r_{AH}} = \left[\frac{\partial m(v,r)}{\partial M(v)}\Big|_{r=r_{AH}}\right] dM(v) = T_{AH} dS_{AH} + \Phi_{AH}^{\alpha} d\alpha + \Phi_{AH}^{\beta} d\beta + \cdots. \tag{66}$$

At the apparent horizon, $T_{AH}$ and $\tilde{T}_{AH}$ are also related by the corrected factor

$$T_{AH} = \left[\frac{\partial m(v,r)}{\partial M(v)}\Big|_{r=r_{AH}}\right] \tilde{T}_{AH}. \tag{67}$$

For the dynamical Hayward black hole and the dynamical non-commutative black hole,

the correction factors are given by, respectively

$$F(v,r_{AH},\alpha) = \frac{\partial m(v,r,\alpha)}{\partial M(v)}\bigg|_{r=r_{AH}} = \left[\frac{\partial}{\partial M(v)}\left(\frac{M(v)r^3}{r^3 + 2\alpha M(v)}\right)\right]\bigg|_{r=r_{AH}} = \frac{1}{r_{AH}^4}\left(r_{AH}^2 - \alpha\right)^2,$$
(68)

$$F(v,r_{AH},\vartheta) = \frac{\partial m(v,r,\vartheta)}{\partial M(v)}\bigg|_{r=r_h} = \left\{\frac{\partial}{\partial M(v)}\left[\frac{2M(v)}{\sqrt{\pi}}\gamma\left(\frac{3}{2},\frac{r^2}{4\vartheta}\right)\right]\right\}\bigg|_{r=r_{AH}} = \frac{2}{\sqrt{\pi}}\gamma\left(\frac{3}{2},\frac{r_{AH}^2}{4\vartheta}\right).$$
(69)

The product of $T_{AH}$ and the correction factor exactly gives $\tilde{T}_{AH}$, which satisfies Eq. (67). This verifies the validity of the corrected first law of thermodynamics at the apparent horizon.

**5.2 Some remarks**

In this subsection, we provide some discussion regarding the corrected first law of thermodynamics. From the differential form of the field equation (40.2) or its integral form Eq. (42), it is not difficult to find that the $m(v,r)$ appearing in the correction factor is given by

$$m(v,r) = -4\pi \int_0^r r^2 T_v^v dr.$$
(70)

Its form suggests that it is likely related to the energy enclosed within a sphere of radius $r$. In fact, we can verify that the Misner-Sharp energy within a sphere of radius of $r$ is exactly equal to $m(v,r)$. This can be verified using the definition of Misner-Sharp energy

$$E_{MS}(v,r) = \frac{r}{2}\left(1 - h^{ab}\partial_a r \partial_b r\right) = \frac{r}{2}[1 - f(v,r)] = m(v,r).$$
(71)

Therefore, the corrected first law of thermodynamics for spherically symmetric dynamical regular black hole can be written as

$$\left[\frac{\partial E_{MS}(v,r)}{\partial M(v)}\bigg|_{r=r_{(A)H}}\right] dM(v) = T_{r_{(A)H}} dS_{r_{(A)H}} + \Phi^\alpha_{(A)H} d\alpha + \Phi^\beta_{(A)H} d\beta + \cdots.$$
(72)

For static spherically symmetric black holes, the correction factor can also be expressed in the same form. Here, the reason we choose the Misner-Sharp energy is that it is a conserved charge associated with the Kodama vector. However, a more physically intuitive explanation for why the corrected factor of spherically symmetric regular black holes is related to the Misner-Sharp energy still remains to be explored.

Additionally, by examining the forms of Eqs. (50) and (66), we can see that in general $d\mathcal{M}(v)$ is not an exact differential form. Only when $\partial T_v^v / \partial M(v) = 0$, that is, when $d\mathcal{M}(v) = dM(v)$, does it become an exact differential form. In this way, the corrected first law of thermodynamics reduces to the traditional first law of thermodynamics for the singular black holes.

Another issue worth discussing is that, since we can successfully establish a corrected first law of thermodynamics on both the event horizon and the apparent horizon, which one should be considered more fundamental? It is noticed that, Refs. [23,24] treat the event horizon as a time-dependent perturbation of the apparent horizon, and show that thermodynamics can also be established on the event horizon. Therefore, the authors conclude that the thermodynamics associated with the apparent horizon should be regarded as more fundamental. However, if we think in reverse, the apparent horizon can also be viewed as a time-dependent perturbation of the event horizon. Hence, the explanation given in [23,24] is more of a mathematical interpretation. However, regarding this issue, we believe that the following two points may be particularly important. The first concerns the definition of the event horizon itself, which can only be defined in asymptotically flat space-times. However, the universe may not be asymptotically flat, so the existence of an event horizon is questionable. Even if the universe were asymptotically flat, an observer with a finite lifespan would not be able to verify such a global property [33]. The second point is related to Ref. [21], in which the author considered the collapse process of a spherical shell and suggested that the Hawking effect is associated with the apparent horizon rather than the event horizon. The reason was that the apparent horizon constitutes the boundary of the ergoregion, and if Hawking radiation originates from regions near the ergoregion, then the apparent horizon is associated with Hawking radiation. Another conclusion reached in [21] is even more significant: Hawking radiation survives even in the absence of an event horizon. Therefore, in this paper, we are more inclined to believe that the corrected first law of thermodynamics on the apparent horizon is more fundamental.

## 6 Summary

Inspired by the corrected first law of thermodynamics for static spherically symmetric regular black holes, we investigate the corrected first law of thermodynamics for dynamical regular black holes. First, we calculate the apparent horizon radius and the event horizon radius for an arbitrary spherically symmetric dynamical regular black hole, and find that they are very close to each other. Then, for the event horizon and the apparent horizon, we obtain the corresponding temperatures using the conformal flat method and the surface gravity method, respectively. However, the temperature derived from the first law of thermodynamics differs from those obtained by the two methods mentioned above. This indicates that, similar to the static case, the first law of thermodynamics also needs to be corrected for dynamical regular black holes. Following the idea in [23,24], that thermodynamics can be built on both horizons, we successfully establish the corrected first law of thermodynamics on both horizons using the Einstein field equations. Moreover, we find that they share the same form. Finally, we find that for both spherically symmetric dynamical and static regular black holes, the correction factor is related to the derivative of the Misner-Sharp energy with respect to the mass $M(v)$. The deeper physical meaning behind this relation requires further investigation.

## Appendix

In this appendix, we explicitly calculate the apparent horizon radius for the dynamical black hole described by the line element (10). First, for computational convenience, we rewrite it in a metric signature of (+, −, −, −) as

$$ds^2 = f(v,r)dv^2 - 2dvdr - r^2d\theta^2 - r^2\sin^2\theta d\varphi^2. \tag{A.1}$$

The determinant and non-vanishing contravariant components of the metric are given by, respectively,

$$g = -r^4 \sin^2 \theta,$$

$$g^{vr} = g^{rv} = -1, \quad g^{rr} = -f(v,r), \quad g^{\theta\theta} = -\frac{1}{r^2}, \quad g^{\varphi\varphi} = -\frac{1}{r^2 \sin^2 \theta}. \tag{A.2}$$

From line element (A.1), we can obtain the non-vanishing components of the affine connections as

$$\Gamma^v_{vv} = \frac{1}{2}f', \quad \Gamma^v_{\theta\theta} = -r, \quad \Gamma^v_{\varphi\varphi} = -r\sin^2 \theta,$$

$$\Gamma^r_{vr} = \Gamma^r_{rv} = -\frac{1}{2}f', \quad \Gamma^r_{vv} = \frac{1}{2}\left(-\dot{f} + ff'\right),$$

$$\Gamma^r_{\theta\theta} = -f \cdot r, \quad \Gamma^r_{\varphi\varphi} = -f \cdot r \sin^2 \theta, \tag{A.3}$$

$$\Gamma^\theta_{r\theta} = \Gamma^\theta_{\theta r} = \frac{1}{r}, \quad \Gamma^\theta_{\varphi\varphi} = -\sin\theta\cos\theta,$$

$$\Gamma^\varphi_{r\varphi} = \Gamma^\varphi_{\varphi r} = \frac{1}{r}, \quad \Gamma^\varphi_{\theta\varphi} = \Gamma^\varphi_{\varphi\theta} = \cot\theta.$$

Next, we choose the following null tetrad

$$l_\mu = \left(\frac{f}{2}, -1, 0, 0\right), \quad n_\mu = (1, 0, 0, 0),$$

$$m_\mu = \frac{r}{\sqrt{2}}(0, 0, 1, i\sin\theta), \tag{A.4}$$

$$\bar{m}_\mu = \frac{r}{\sqrt{2}}(0, 0, 1, -i\sin\theta),$$

and their contravariant forms are given by

$$l^\mu = \left(1, \frac{f}{2}, 0, 0\right), \quad n^\mu = (0, -1, 0, 0),$$

$$m^\mu = \frac{1}{\sqrt{2}r}\left(0, 0, -1, -\frac{i}{\sin\theta}\right), \tag{A.5}$$

$$\bar{m}^\mu = \frac{1}{\sqrt{2}r}\left(0, 0, -1, \frac{i}{\sin\theta}\right).$$

One can verify that they satisfy the condition of the null tetrad frame

$$n_\mu n^\mu = l_\mu l^\mu = m_\mu m^\mu = \bar{m}_\mu \bar{m}^\mu = 0,$$

$$n_\mu l^\mu = -m_\mu \bar{m}^\mu = 1, \tag{A.6}$$

$$n_\mu m^\mu = n_\mu \bar{m}^\mu = l_\mu m^\mu = l_\mu \bar{m}^\mu = 0.$$

Using the components of the affine connection and the null tetrads, we can calculate $l^\mu_{;\mu}$ and $\kappa$ as follows

$$l^{\mu}_{;\mu} = l^{\mu}_{,\mu} + \Gamma^{\mu}_{\sigma\mu}l^{\sigma} = \frac{1}{\sqrt{-g}}\frac{\partial}{\partial x^{\mu}}\left(\sqrt{-g}l^{\mu}\right) = \frac{1}{\sqrt{-g}}\left[\frac{\partial}{\partial v}\left(\sqrt{-g}l^{v}\right) + \frac{\partial}{\partial r}\left(\sqrt{-g}l^{r}\right)\right] = \frac{rf' + 2f}{2r},$$

(A.7)

$$\kappa = \left(l_{\mu,\nu} - \Gamma^{\sigma}_{\mu\nu}l_{\sigma}\right)n^{\mu}l^{\nu} = l_{\mu,\nu}n^{\mu}l^{\nu} - \Gamma^{\sigma}_{\mu\nu}l_{\sigma}n^{\mu}l^{\nu} = \left(l_{r,v}n^{r}l^{v} + l_{r,r}n^{r}l^{r}\right) - \Gamma^{r}_{rv}l_{r}n^{r}l^{v} = \frac{f'}{2}.$$

(A.8)

Therefore, we can obtain $\Theta = l^{\mu}_{;\mu} - \kappa = f(v,r)/r$. Setting $\Theta = 0$, we finally obtain that the radius of the apparent horizon satisfies $f(v, r_{AH}) = 0$.